\documentclass[iop]{emulateapj} 

\usepackage{amssymb}
\usepackage{multirow}

\def\msun{$M_{\odot}$}

\def\ergsec{\hbox{erg s$^{-1}$}} 
\def\ergcm{\hbox{erg s$^{-1}$ cm$^{-2}$}}

\def\xmm{\emph{XMM-Newton}}
\def\chandra{\emph{Chandra}}
\def\swift{\emph{Swift}}

\shorttitle{State transitions of ESO 243--49 HLX-1}
\shortauthors{Servillat et al.}

\begin{document}

\title{X-ray Variability and Hardness of ESO 243--49 HLX-1: \\Clear Evidence for Spectral State Transitions}

\author{Mathieu Servillat\altaffilmark{1}, Sean A. Farrell\altaffilmark{2,3}, Dacheng Lin\altaffilmark{4,5}, Olivier Godet\altaffilmark{4,5}, \\Didier Barret\altaffilmark{4,5}, Natalie A. Webb\altaffilmark{4,5}}

\affil{\altaffilmark{1}Harvard-Smithsonian Center for Astrophysics, 60 Garden Street, MS-67, Cambridge,~MA~02138 -- {mservillat@cfa.harvard.edu}}
\affil{\altaffilmark{2}Department of Physics and Astronomy, University of Leicester,
University Road, Leicester, LE1 7RH, UK}
\affil{\altaffilmark{3}Sydney Institute for Astronomy, School of Physics A29, The
University of Sydney, NSW 2006, Australia}
\affil{\altaffilmark{4}Universit\'e de Toulouse; Universit\'e Paul Sabatier -- Observatoire
Midi-Pyr\'en\'ees, \\Institut de Recherche en Astrophysique et Plan\'etologie
(IRAP), Toulouse, France}
\affil{\altaffilmark{5}Centre National de la Recherche Scientifique; IRAP; 9 Avenue
du colonel Roche, BP 44346, F-31028 Toulouse cedex 4, France}


\begin{abstract}
The ultra-luminous X-ray (ULX) source ESO 243--49 HLX-1, which reaches a maximum luminosity of $10^{42}$~\ergsec\ (0.2--10~keV), currently provides the strongest evidence for the existence of intermediate mass black holes. 
To study the spectral variability of the source, we conduct an ongoing monitoring campaign with the \swift\ X-ray Telescope, which now spans more than two years. 
We found that HLX-1 showed two fast rise and exponential decay (FRED) type outbursts in the \swift\ XRT light-curve with increases in the count rate of a factor $\sim$40 separated by 375$\pm$13 days.
We obtained new \xmm\ and \chandra\ dedicated pointings that were triggered at the lowest and highest luminosities, respectively. From spectral fitting, the unabsorbed luminosities ranged from $1.9\times10^{40}$ to $1.25\times10^{42}$~\ergsec.
We confirm here the detection of spectral state transitions from HLX-1 reminiscent of Galactic black hole binaries:
at high luminosities, the X-ray spectrum showed a thermal state dominated by a disk component with temperatures of 0.26~keV at most, and at low luminosities the spectrum is dominated by a hard power law with a photon index in the range 1.4--2.1, consistent with a hard state.
The source was also observed in a state consistent with the steep power law state, with a photon index of $\sim3.5$.
In the thermal state, the luminosity of the disk component appears to scale with the fourth power of the inner disk temperature which supports the presence of an optically thick, geometrically thin accretion disk. The low fractional variability (rms of $9\pm9$\%) in this state also suggests the presence of a dominant disk. 
The spectral changes and long-term variability of the source cannot be explained by variations of the beaming angle and are not consistent with the source being in a super-Eddington accretion state as is proposed for most ULX sources with lower luminosities.
All this indicates that HLX-1 is an unusual ULX as it is similar to Galactic black hole binaries, which have non-beamed and sub-Eddington emission, but with luminosities 3 orders of magnitude higher.
In this picture, a lower limit on the mass of the black hole of $>$9000 $M_\odot$ can be derived, and the relatively low disk temperature in the thermal state also suggests the presence of an intermediate mass black hole of a few $10^{3}$~\msun.
\end{abstract}


%

\section{Introduction}


At present, only two families of black holes have convincing observational evidence: stellar-mass black holes detected in some X-ray binaries \cite[e.g.][]{Remillard:2006p561} and super-massive black holes (SMBHs, 10$^{6-9}$~\msun) which are ubiquitous in the center of galaxies \citep{Kormendy:1995p1434}, sometimes revealing themselves as active galactic nuclei (AGN). Remarkably, the existence of intermediate mass black holes (IMBHs, 10$^{2-5}$~\msun) remains to be proven. Such objects may have been ejected following an interaction with the central SMBH and may remain in the halo of galaxies \citep{Micic:2011p1896}. They may be found in globular clusters, and the strongest case is the massive cluster G1 in M31, where dynamical, X-ray, and radio studies are consistent with a mass of $2\times10^{4}$~\msun\ \citep{Gebhardt:2002p3007,Kong:2010p5289}. However, radio observations of Galactic globular clusters currently provide only upper limits on the mass \cite[e.g.][]{Maccarone:2008p2238,Lu:2011p2630}.
IMBHs may also be the nuclei of satellite galaxies captured during hierarchical merging \citep{King:2005p5167}. Finally, they may be the engine of some ultra-luminous X-ray sources \citep[ULXs, e.g.][]{Miller:2004p1567}. 

A ULX is a non-nuclear extragalactic X-ray source that has an X-ray luminosity that exceeds the Eddington luminosity --- $1.3\times10^{38}~(M_\mathrm{BH}/M_{\odot})$~\ergsec\ --- for a stellar-mass black hole ($\sim$10~\msun), supposing isotropic emission \citep[see][for a review]{Roberts:2007p3209}. 
Different solutions have been proposed for the high luminosity problem of ULXs: (1) they are stellar-mass black holes emitting up to a factor of 10 above their Eddington limit \citep{Begelman:2002p4262} in a new ''ultra-luminous'' state with super-Eddington accretion rates \citep[e.g.][]{Gladstone:2009p2047}, (2) they are stellar-mass black holes with geometric \citep[e.g.][]{King:2009p2048} or relativistic \citep{Kording:2002p5290} beaming,
or (3) the objects are IMBHs \citep[e.g.][]{Miller:2004p1567}. It could also be a combination of these possibilities. For the most luminous ULXs above $10^{41}$~\ergsec, the mass argument seems to be the main one that can convincingly explain the extreme luminosities. 


One of the keys to further constraining the nature of the black hole in such sources is the study of their spectral and timing variability.
Indeed, Galactic black hole binaries (GBHBs) containing stellar-mass black holes have been commonly observed to undergo transitions between different spectral states (e.g. \citealt{Tananbaum:1972p1616,Kubota:2004p5432}, see \citealt{Remillard:2006p561,Done:2007p1217,Belloni:2010p1608} for reviews).
In the thermal state, the emission is dominated by an optically thick, geometrically thin accretion disk component \citep{Shakura:1973p1981}, with temperatures of $\sim$1~keV \citep[e.g.][]{Remillard:2006p561}. {For such a disk extending down to the innermost stable circular orbit (ISCO), the luminosity $L_{\rm disk}$, inner temperature $T_{\rm in}$ and mass of the black hole $M_{\rm BH}$ are theoretically related: $L_{\rm disk} \propto T^{4}_{\rm in}$ for a given $M_{\rm BH}$, and $T_{\rm in} \propto M^{-1/4}_{\rm BH}$ \citep{Shakura:1973p1981,Makishima:2000p5954}.}
In the hard state, the accretion disk appears to be fainter and cooler, and may be truncated at a large radius. The physical condition of material within this radius remains uncertain. Investigations of GBHBs in the hard state suggest that both synchrotron and Compton components contribute to the broadband, power law like spectrum with typical photon indices of $1.4 < \Gamma < 2.1$ \citep[e.g.][]{Remillard:2006p561}. Quasi-steady radio jets are commonly observed during this state, and clear correlations between the radio and X-ray intensities have been reported \citep{Fender:2004p3226}.
Some GBHBs also showed a steep power law state \citep[$\Gamma > 2.4$, e.g.][]{Remillard:2006p561}. 

Consideration of simple properties such as X-ray spectral hardness and fractional variability have led to a better understanding of the physical processes in action \citep[e.g.][]{Belloni:2010p1608}. In a hardness-intensity diagram (HID), the source shows a hysteresis curve between the thermal state and the hard state \citep{Miyamoto:1995p5433,Maccarone:2003p5296}. The root mean square (rms) variability amplitude is generally low in the thermal state ($<$10\%), and can reach higher values in the hard state, with typical rms of $\sim$30\%. As seen in a hardness-rms diagram (HRD), the rms is anti-correlated with flux and positively correlated with hardness \citep{Belloni:2010p1608}. 

{The lack of a canonical thermal state, with a dominant disk component, seems to be a common feature of ULXs \citep{Feng:2005p6042,Winter:2006p6226,Soria:2009p2406}.
Two ULXs showed such a thermally dominated spectrum: M82 X41.4+60=X-1 \citep[$kT_{in}=1.1$--$1.5$~keV and $L_{\rm X}=2$--$8\times10^{40}$~\ergsec,][]{Feng:2010p2397}, and M82 X37.8+54 \citep[$0.9$--$1.6$~keV and $L_{\rm Xmax}=4.4\times10^{39}$~\ergsec,][]{Jin:2010p2378} . Those sources could harbor a fast spinning black hole of 200--800~\msun\ and $<$100~\msun, respectively.
ULXs with $kT_{\rm in} \sim 0.1$~keV soft disk spectra may be good candidates for IMBHs in the thermal state. Some ULXs have shown soft excesses that could be fitted by such a cool disk emission \cite[e.g.][]{Miller:2003p1982}. However, this component is not dominant, thus the sources are not in the canonical thermal state.
A few intriguing ultraluminous supersoft sources show temperatures of $\sim$0.1~keV \citep[e.g.][]{Fabbiano:2003p6428,Mukai:2003p6429,Kong:2004p6320,Jin:2011p2421}, but they could be explained by matter outflow from super-Eddington accretion \citep{King:2003p6426}.}




ULXs do not generally show qualitative spectral changes when the luminosity varies \citep{Feng:2005p6042,Kaaret:2009p2046,Grise:2010p2044}.
Possible state transitions have been reported for some ULXs but they are subtle and not clearly similar to GBHBs: e.g. Holmberg IX X-1 \citep{LaParola:2001p1993,Vierdayanti:2010p2493} or M82 X37.8+54 \citep{Jin:2010p2378}.

ULXs are generally highly variable sources showing variations of fluxes by a factor of $\sim$5--15 on time scales of hours to years \citep{LaParola:2001p1993,Grise:2010p2044,Kong:2011p5036}.
On shorter time scales, six ULXs showed intrinsic variability with power spectra in the form of either a power-law or broken-power-law-like continuum \citep{Heil:2009p1986} and in some cases quasi-periodic oscillations (QPOs, \citealt{Casella:2008p5323,Strohmayer:2009p2453}). In some ULXs the variability is significantly suppressed compared to bright GBHBs and AGN \citep{Heil:2009p1986}.



To date, one of the best IMBH candidates is the most luminous ULX ESO 243--49 HLX-1 (HLX-1 for short) which reaches a luminosity of $10^{42}$~\ergsec\ \citep{Farrell:2009p1164}. 
The distance, and thus the high X-ray luminosity of HLX-1, have been firmly confirmed through the detection of an H$\alpha$ line at a redshift consistent with that of its host galaxy \citep{Wiersema:2010p1170}. The H$\alpha$ line has been found in the spectrum of the optical counterpart which falls inside the \chandra\ error circle of the source ($0\farcs3$ at a 95\% confidence level, \citealt{Webb:2010p1169}; see also \citealt{Soria:2010p1603}).
Observations of HLX-1 with the \swift\ X-ray Telescope (XRT) in August 2009 showed a hardening of the source at low luminosities which was interpreted as the first evidence for a change to the hard spectral state \citep{Godet:2009p1106}.

Our aim in this paper is to measure the spectral hardness and the variability of HLX-1 at different luminosity states, using all the X-ray data available from current observatories\footnote{Except two \chandra\ HRC-I observations \citep{Webb:2010p1169} as this instrument does not provide spectral information.} in order to test similarities with GBHBs or other ULXs.
For this purpose, we obtained new \xmm\ and \chandra\ dedicated pointings that were triggered at the lowest and highest luminosities, respectively.
The X-ray datasets are described in Section~\ref{data}. 
Spectral analysis of firstly the high luminosity states (Section~\ref{high}) and then the low luminosity state (Section~\ref{low}) follows.
A global analysis is given in Section~\ref{diagrams} and the final Section~\ref{discuss} discusses the properties of HLX-1 compared to GBHBs and other ULXs.

\section{Data and long term variability}
\label{data}

\begin{figure*}[ht]
\centering
\includegraphics[width=\textwidth]{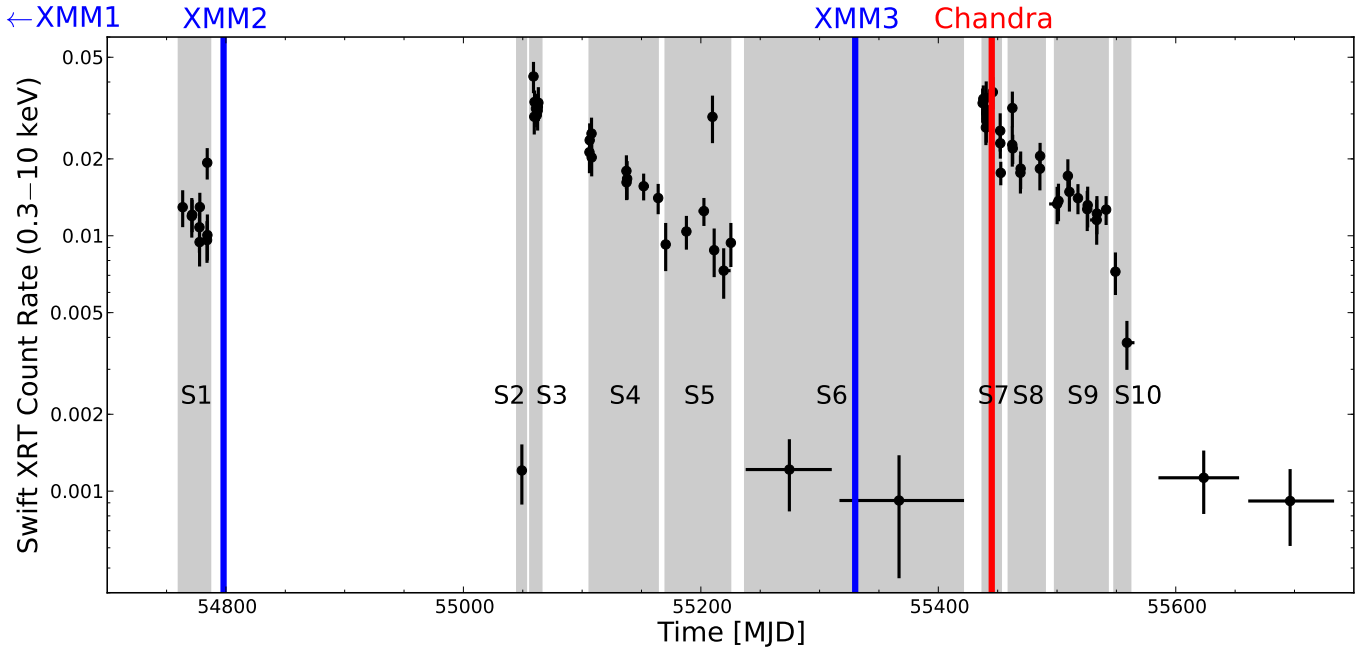}
\caption{\swift\ XRT light curve of HLX-1. The epochs that were grouped to generate spectra are shaded with labels S1 to S10. The dates of the \xmm\ and \chandra\ observations are represented by vertical lines.}
\label{fig:swift_lc}
\end{figure*}

\begin{deluxetable*}{ccccccccc}
\tablecaption{\xmm\ observations of HLX-1 \label{tbl:xmmobs}}
\tablewidth{0pt}
\tablehead{\colhead{Obs. name} & \colhead{ObsID} & \colhead{Date} & \colhead{MJD} & \colhead{Exp. time [ks]} & \colhead{Instrument} & \colhead{Mode} & \colhead{Source region} & \colhead{Background region}}
\startdata
XMM1 & 0560180901 & 2004-09-23 & 53271 & 21 & MOS1\&2, pn & full-frame & 22.5\arcsec & 50.0\arcsec--63.4\arcsec \\
[0.5ex] \tableline \\[-2ex]
XMM2 & 0560180901 & 2008-11-28 & 54798 & 50 & MOS1\&2 & full-frame & 27\arcsec & 60\arcsec--76\arcsec \\
           &                  &                  &          &  & pn & small window & 27\arcsec & 3$\times$27\arcsec \\
[0.5ex] \tableline \\[-2ex]
XMM3 & 0655510201 & 2010-05-14 & 55330 & 100 & MOS1\&2 & full-frame & 14\arcsec & 40\arcsec--47\arcsec \\
           &                  &                  &          &  & pn & small window & 14\arcsec & 3$\times$14\arcsec
\enddata 
\end{deluxetable*}



\subsection{\swift\ XRT data}


HLX-1 has been regularly monitored by the \swift\ XRT since October 2008.
Since August 2009, observations occurred on average every $\sim$1--2 weeks.
All the data were processed with the \swift\ XRT pipeline\footnote{http://heasarc.gsfc.nasa.gov/docs/swift/analysis/} version 0.12.4. We used the grade 0-12 events, giving slightly higher effective area at higher energies than grade 0 events. 
The background extraction region, chosen to be close to the source extraction region, is the same for all epochs. No \xmm\ sources are present inside the background extraction region.

We generated a light curve from all the \swift\ pointings with a binning of a minimum of 50 counts
per bin for the energy band 0.3-10 keV (Figure~\ref{fig:swift_lc}) using the web interface \citep{Evans:2007p6460,Evans:2009p5075}.
There is a recurrence in the light curve of 375$\pm$13~days between the two increases that may be periodic, but this will have to be confirmed by further observations. 
If the behavior is indeed periodic, the next maximum would occur in August/September 2011.
The cause of this variability has been investigated in more details by \citet{Lasota:2011p4895}.

We combined the data into different epochs labelled S1 to S10 in Figure~\ref{fig:swift_lc}, and extracted a spectrum for each epoch. A 20-pixels ($47\farcs2$) radius circle was used to extract the source and the background spectra using \textsc{XSELECT} v2.4a. The ancillary files were created with \textsc{XRTMKARF} v0.5.6 and exposure maps generated with \textsc{XRTEXPOMAP} v0.2.7. The response file swxpc0to12s6\_20070901v011.rmf was used in the spectral fitting process. We used a binning of a minimum of 20 counts per bin for each spectrum except the S2 and S6 spectra in order to use the $\chi^{2}$ statistic within Xspec 12.6 \citep{Arnaud:1996p4268}. For the S2 and S6 spectra, we used the Cash statistic \citep{Cash:1979p5198} due to the lower number of counts in these spectra.

\subsection{\xmm\ EPIC data}


The field of ESO 243--49 and HLX-1 was observed three times with \xmm\ (Table~\ref{tbl:xmmobs}). 
HLX-1 was detected serendipitously at an off-axis angle of $9\farcm29$ in XMM1, and XMM2 was obtained to show the spectral variability of HLX-1 \citep{Farrell:2009p1164}. Following a transition to a very low flux state, we triggered the XMM3 observation.


The data were processed using the \xmm\ Science Analysis System\footnote{http://xmm.esa.int/sas/} (SAS) v10.0 software with the most recent calibration files as of 2010 December 13. The observation data files (ODF) were reduced using the \emph{epproc} and \emph{emproc} tasks to produce event lists. Single event light curves with energies exceeding 10~keV were generated for each camera in each observation in order to identify periods of high background related to soft proton flares. In the XMM1 and XMM2 observations the background levels were low in each camera with no flaring events, and so no good time interval (GTI) filtering was applied. However, during the XMM3 observation significant flaring events were present at the start and end of the exposure, with an additional small short-duration flare occurring $\sim$16~ks into the observation. For the XMM3 data we therefore 
generated GTI files for the spectral extraction using the SAS task \emph{tabgtigen} and cut-off count rates of 0.3~counts~s$^{-1}$ and 0.1~counts~s$^{-1}$ in the flare background light curves for the MOS and pn cameras respectively. This filtering resulted in net exposure times of $\sim$98~ks and $\sim$99~ks (for pn and MOS respectively).
The same event filtering criteria as used for the production of pipeline products for the 2XMM catalogue were used to produce light curves and spectra for each camera in each observation (e.g. single to double events for the pn and single to quadruple events for the MOS, \citealt{Watson:2009p1813}).

Source and background spectra were extracted for each camera for each observation, with response and ancillary files generated in turn using the tasks \emph{rmfgen} and \emph{arfgen}. 
The source extraction region radii (reported in Table~\ref{tbl:xmmobs}) were chosen so as to optimize the signal-to-noise based on the detected count rates and off-axis position of HLX-1 in each of the observations (V.~Braito, private communication).
The background extraction regions were chosen to be three times the area of the source extraction regions, so as to provide a robust estimate of the average background level at the position of HLX-1.
The spectra were grouped to have at least 20 counts per bin to provide sufficient statistics for spectral analyses using $\chi^2$ statistic.

Source light curves were extracted using events in the energy range of 0.3--2 keV for the pn camera of all observations. They were binned at the frame time, which is 73.4 ms for XMM1 and is 5.9 ms for XMM2 and XMM3.

\begin{figure}
\centering
\includegraphics[width=\columnwidth]{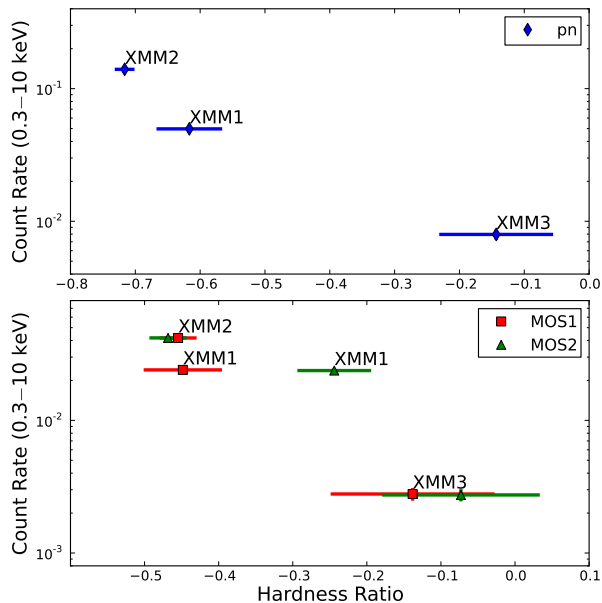}\\
\caption{Hardness intensity diagram for HLX-1 based on \xmm\ count rates (counts s$^{-1}$). Hardness ratios are HR = (B2$-$B1) / (B1+B2) with B1: 0.3$-$1 keV and B2: 1$-$10 keV. The discrepency between MOS1 and 2 in XMM-1 was already identified in Farrell  et al. (2009).}
\label{fig:xray_hid_xmm}
\end{figure}

Figure~\ref{fig:xray_hid_xmm} shows the count rates for each camera as a function of a hardness ratio. 
We see a discrepancy between the XMM1 MOS1 and MOS2 hardness ratios,
with the MOS2 data showing a deficit in counts in the 0.3--1 keV energy band
compared to both the MOS1 and pn data\footnote{This discrepancy had been
noted earlier by Farrell et al. (2009), leading them to exclude energies
below 0.5 keV for MOS2.}. We thus excluded energies below 0.5 keV for MOS2. Despite this inconsistency between the MOS hardness ratios
in the XMM1 data, a clear trend is observed indicating significant spectral
hardening at lower luminosities, which is most clearly shown by the pn data
(Figure 2, top panel).

\subsection{\chandra\ ACIS-S data}

We obtained a 10~ks observation with the AXAF CCD Imaging Spectrometer (ACIS) onboard \chandra\ on 2010 September 6 (ObsID 13122, MJD 55445) in the director's discretionary time program after the source showed a re-brightening in X-rays \citep{Godet:2010p4890}.
In order to limit pile-up, this observation was performed with the ACIS-S3 chip only and using a 1/4 chip subarray with a frame time of 0.8 s, leading to 5\% dead time (0.041 s). The ACIS-S3 chip was chosen for its higher sensitivity to softer X-ray photons\footnote{http://cxc.harvard.edu/proposer/POG, Proposers' Observatory Guide}.

The data were processed using CIAO\footnote{http://cxc.harvard.edu/ciao/} 4.2 and CALDB 4.3.0 \citep{Fruscione:2006p4909}.
We recreated the evt2 file following the corresponding thread\footnote{http://cxc.harvard.edu/ciao/threads/createL2} to apply the latest calibration files. We extracted the source spectrum using \textit{psextract} with an extraction circle of radius $2\farcs2$ which includes 90\% of the energy at 8~keV. The background spectrum was extracted in an annulus surrounding the source with inner and outer radii of 15\arcsec\ and 50\arcsec\ respectively (thus excluding the host galaxy). We then ran the tasks \textit{mkacisrmf} and \textit{mkarf} to obtain the instrument response files as recommended for ACIS-S data. The spectrum was binned with a minimum of 15 counts, and we discarded bins below 0.3~keV where the response of the instrument is not calibrated.

\section{HLX-1 at high luminosities}
\label{high}

\begin{deluxetable*}{@{~~~~}cccccccc}
\tablecaption{Parameters of best fits for HLX-1 X-ray spectra \label{tbl:specfits}}
\tablewidth{0pt}
\tablehead{\colhead{Obs} & \colhead{$N_{\rm H20}$} & \colhead{$kT_{\rm in}$} & \colhead{$\Gamma$} & \colhead{$\chi^{2}$/dof} & \colhead{$F_{\rm [0.2-10~keV]}$} & \colhead{unabs. $L_{\rm [0.2-10~keV]}$} & \colhead{unabs. $L_{\rm disk}$} \\
 (1) & (2) & (3) & (4) & (5) & (6) & (7) & (8) 
}
\startdata
\multicolumn{8}{l}{Steep power law state} \\[0.5ex] 
XMM1   & $10\pm4$  & \nodata  & $3.5^{+0.3}_{-0.2}$  & 59.7/64  & $3.2\pm0.5$ & $9.9\pm0.5$ & \nodata  \\
XMM1   & $5\pm5$    & $0.27^{+0.12}_{-0.05}$  & $3.2^{+0.3}_{-0.4}$  & 56.3/62  & $3.1\pm1.2$ & $6.9\pm2.6$ & $1.6^{+0.7}_{-1.2}$  \\
[0.5ex] \tableline \\[-2ex]
\multicolumn{8}{l}{Thermal state} \\[0.5ex] 
\chandra\tablenotemark{a} & $2^{+4}_{-2}$  & $0.22\pm0.02$ & \nodata & 57.0/48  & $6.7\pm0.3$  & $9.0\pm0.3$ & $13.0\pm0.3$ \\
XMM2        & $3\pm1$  & $0.18\pm0.01$  & $2.1\pm0.2$  & 395.6/346  & $3.5\pm0.2$  & $6.4\pm0.2$ & $6.6\pm0.3$  \\
S1+S4+S8 & 3  & $0.18\pm0.02$  & $2.3^{+0.7}_{-0.5}$  &  48/45 & \nodata & \nodata & \nodata \\
S3 & 3       & $0.26\pm0.03$  & \nodata  & 14.1/21  &  $8.1\pm0.5$  & $12.2\pm0.5$   & $15.2\pm0.5$   \\
S7 & 3       & $0.23\pm0.03$  & \nodata  & 12.3/14  & $7.7\pm0.5$  & $12.5\pm0.5$  & $16.5\pm0.5$   \\
[0.5ex] \tableline \\[-2ex]
\multicolumn{8}{l}{Hard state} \\[0.5ex] 
XMM3                          & $6^{+6}_{-4}$  & \nodata  & $2.5^{+0.6}_{-0.4}$  & 85.2/60   & $0.17^{+0.01}_{-0.09}$  & $0.31^{+0.02}_{-0.10}$ & \nodata \\
XMM3                          & $62^{+40}_{-6}$  & $0.09\pm0.02$  & $2.4^{+0.5}_{-0.8}$  & 69.3/58   & $0.17^{+0.02}_{-0.05}$  & $\sim$21  & $\sim$56 \\
XMM3\tablenotemark{b} & 3  & \nodata  & $2.0\pm0.3$  & 63.6/59   & $0.16\pm0.03$  & $0.19\pm0.03$ & \nodata \\
XMM3\tablenotemark{b} & 3  & $0.07\pm0.04$  & $1.6\pm0.4$  & 58.0/57   & $0.20\pm0.04$  & $0.28\pm0.04$& $0.4\pm0.1$ \\
S2+S6                          & 3  & \nodata  & $2.1\pm0.4$  & 8.4/8   & $0.6\pm0.2$  & $0.8\pm0.3$ & \nodata
\enddata 
\tablenotetext{a}{The \chandra\ spectrum was fitted with a pile-up model with a frame time of 0.8 s, alpha=0.99 and psffrac=0.95, which was then removed to estimate fluxes.}
\tablenotetext{b}{We added a \textit{mekal} component at a fixed redshift of $z=0.0224$ and with solar abundances which significantly improved the fit. We found a temperature $kT=0.43\pm0.09$~keV. This component was removed before estimating the flux and luminosity of HLX-1, assuming it is due to the host galaxy.}
\tablecomments{Columns: (1) Observations; (2) Absorption on the line of sight, when no error, it is fixed to the best estimate of $3\pm1\times10^{20}$~atom~cm$^{-2}$ for XMM2; (3) Temperature of the \textit{diskbb} model in keV and 90\% error; (4) Photon index of the power law component and 90\% error; (5) $\chi^{2}$ of the fit and degrees of freedom; (6) Absorbed flux in the 0.2--10 keV energy range for the model, with 1$\sigma$ error in $10^{-13}$~\ergcm; (7) Unabsorbed luminosity in the 0.2--10 keV energy range assuming a distance of 95 Mpc, with 1$\sigma$ error in $10^{41}$~\ergsec; (8) Unabsorbed bolometric luminosity of the disk component assuming a distance of 95 Mpc, with 1$\sigma$ error in $10^{41}$~\ergsec.}
\end{deluxetable*}

\begin{figure}
\centering
\includegraphics[width=\columnwidth]{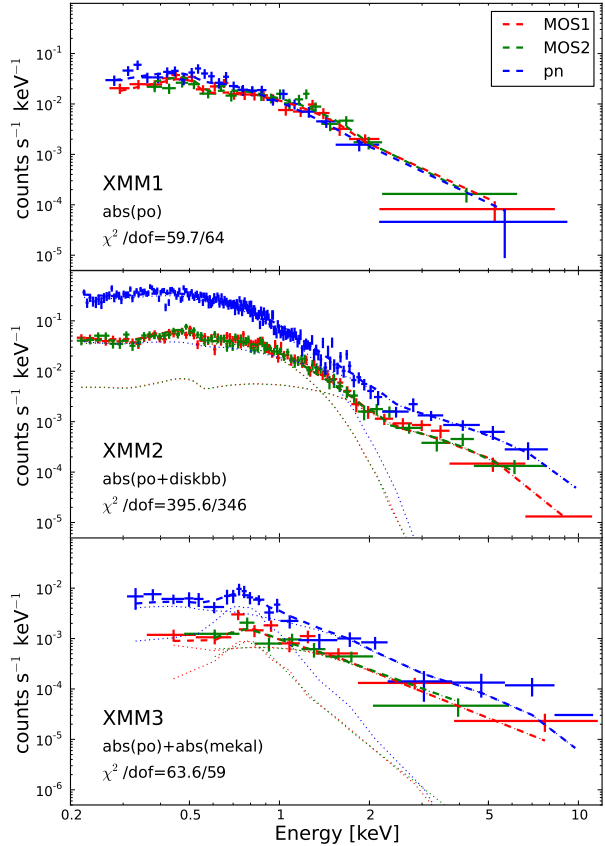}\\
\caption{\xmm\ folded spectra of HLX-1 and best fit models. Data points from the three cameras are reported with their error bars and the model is indicated as a dashed line. Dotted lines show components of the model.}
\label{fig:xray_sp_xmm}
\end{figure}

\subsection{Re-analysis of XMM1 and XMM2 spectra}
\label{xmm12}

Spectral fits of the XMM2 data with the relativistic model BHSPEC have been performed by \citet{Davis:2011p4901}, and a more detailed analysis of all spectra with physically motivated accretion disk models used for GBHBs and ULXs is presented by Godet et al. (2011, in preparation).
We used here simple models to fit the X-ray spectra, e.g. an absorption on the line of sight, a power law model (\textit{pow}), and a multi-temperature black body model (\textit{diskbb}, \citealt{Makishima:1986p1889}). We used the \textit{wabs} absorption model \citep{Morrison:1983p1878}. The Galactic absorption along the line of sight of ESO 243--49 is $1.8\times10^{20}$~atom~cm$^{-2}$ \citep{Kalberla:2005p1892}.


{For XMM1, we fitted simultaneously the MOS1, MOS2 and pn spectra using Xspec and ignored channels with energy lower than 0.3 keV as well as channels tagged as bad. The spectra are well represented by a simple absorbed steep power law of index $3.5^{+0.3}_{-0.2}$ giving a reduced $\chi^{2}$ of 0.93. The addition of a \textit{diskbb} component improved the $\chi^{2}$ by 3.4. To test the significance of this component we used the Bayesian posterior predictive probability values method \citep[e.g.][]{Protassov:2002p2454}, as it was already done by \citet{Farrell:2009p1164}. We generated 3000 simulated set of spectra with the Xspec fakeit command using the absorbed power law model and the response files corresponding to the XMM1 spectra. 
We then computed the distribution of the $\chi^{2}$ improvement of the fit when adding a \textit{diskbb} component. The $\chi^{2}$ improvement of 3.4 for XMM1 translates to a significance of 73\% for the \textit{diskbb} component, insufficient to claim it is real (\citealt{Farrell:2009p1164} derived a significance of 70\% with a similar method).}

{For the XMM2 spectra, we included channels with energy higher than 0.2 keV due to the higher signal to noise at low energies. The spectra are poorly fitted with a simple absorbed power law mode ($\chi^{2}$ of 577.7 and 348 degrees of freedom). When adding a \textit{diskbb} component the reduced $\chi^{2}$ significantly dropped to an acceptable value of 1.14, indicating that this model better represents the spectra.}

For XMM1 and XMM2, we thus obtained best fits in agreement with \citet{Farrell:2009p1164}. We show the corresponding folded spectra in Figure~\ref{fig:xray_sp_xmm}. The calibration was improved in the low energy range with respect to that used by \citet{Farrell:2009p1164} which lead to smaller error bars in this work. We thus present new estimates of the parameters in Table~\ref{tbl:specfits}.

\subsection{\chandra\ spectrum and pile-up}
\label{sec:sp_chandra}

\begin{figure}
\centering
\includegraphics[width=\columnwidth]{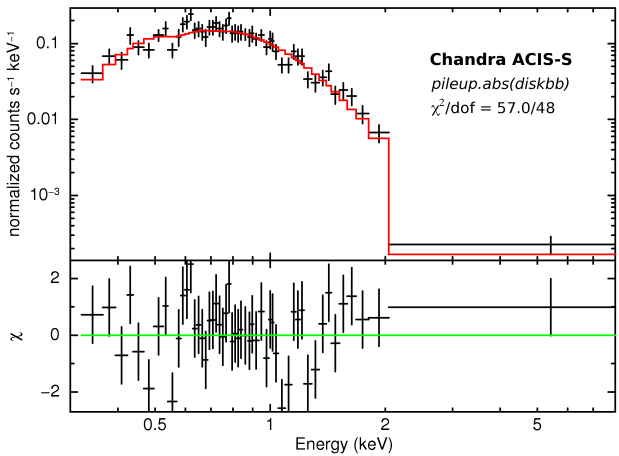}
\caption{\chandra\ ACIS-S folded spectrum of HLX-1. Best fit (top) and residuals (bottom). See Table~\ref{tbl:specfits}.}
\label{fig:ch_sp}
\end{figure}

The spectrum was fitted in Xspec with an absorbed \textit{diskbb} model which gave a reduced $\chi^2$ of 1.43 for 49 degrees of freedom. 
A clear hard excess appeared in the residuals that could correspond to the effect of pile-up \citep{Davis:2001p1847}. 
We indeed estimated a possible level of pile-up of 5\% with the \chandra\ Proposal Planning Toolkit\footnote{http://cxc.harvard.edu/toolkit/pimms.jsp} based on PIMMS v4.2.
We thus added a \textit{pileup} component \citep{Davis:2001p1847} to the model and obtained a satisfactory fit with a single absorbed \textit{diskbb} model with a reduced $\chi^{2}$ of 1.18
(see Figure~\ref{fig:ch_sp} and Table~\ref{tbl:specfits}). We let the parameter alpha free and found a value of $1.0^{+0.0}_{-0.3}$. This parameter is related to the probability of events being retained as a good grade after filtering and we would expect a value between 0.5 and 0.7. 
We note that the value of the \textit{diskbb} temperature is not significantly affected by this parameter. 
Large contributions to the $\chi^{2}$ seem to come from features around 0.6~keV (emission) and 1.1 and 1.3~keV (absorption). However, given the low number of degrees of freedom, there is a large uncertainty on the expectation of the reduced $\chi^{2}$ being 1 \cite[$\sigma$ of 0.2 on the estimate of the reduced $\chi^{2}$, see e.g.][]{Andrae:2010p2114}, so we cannot claim that these features are real. We therefore consider the continuum well fitted by the \textit{diskbb} model.
{We found that an additional power law component with a fixed photon index of 2 would contribute to at most 10\% of the 0.2--10~keV absorbed flux, indicating that the disk component is dominant in the spectrum.}

The \chandra\ observation was simulated to better characterize the emission of HLX-1.
We used the ray tracing tool ChaRT dedicated to \chandra\ \citep{Carter:2003p1368} and generated an image with MARX for the ACIS-S detector \citep{Wise:1997p1369}. The best fit spectral model of the source, obtained as explained above, was used as an input in the energy range 0.3--8~keV. 
A simulated image is presented in Figure~\ref{fig:ch_im}.

In the band 0.3--8 keV, the results from the simulation gave $1294\pm36$ counts. After running the \textit{pileup} tool in MARX, we obtained $1182\pm34$ expected counts ($9\pm5$\% loss). 
We detect $1130\pm34$ counts ($14\pm5$\% loss), which is consistent within the errors with the 
MARX simulation with pile-up effects.
The PSF in the observation shows a deficit of counts in the central pixels compared to the simulation without pile-up. It is however consistent with the image of the simulation with pile-up applied. 
We are therefore confident that the use of the pile-up model in our spectral fitting is justified.

\subsection{Emission from the galaxy}
\label{sec:contamination}

\setlength\fboxsep{0pt}
\setlength\fboxrule{0.8pt}

\begin{figure}
\centering
\fbox{\includegraphics[width=\columnwidth]{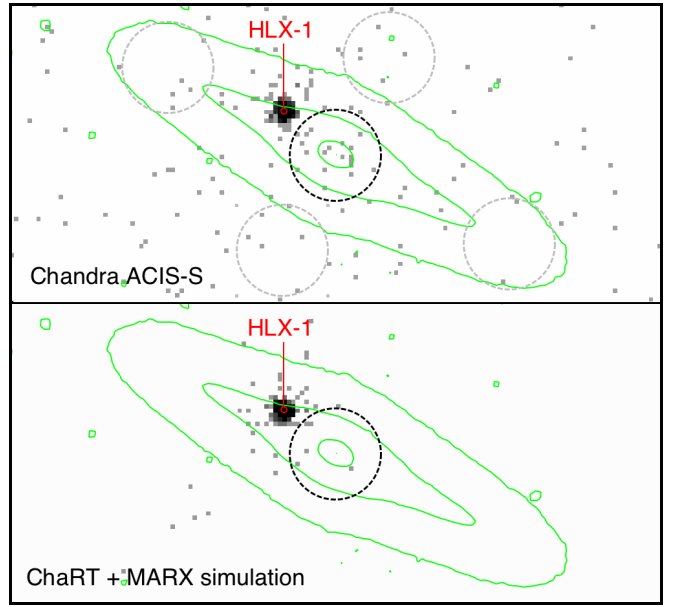}}
\caption{\chandra\ ACIS-S images of HLX-1 in the 0.3-8 keV energy band. \textit{Top}: data. \textit{Bottom}: result of simulation using ChaRT and MARX with pile-up included. Galaxy contours are overlaid (at H-band Vega mag arcsec$^{-1}$ of 21, 18.5 and 16). A dashed black 5\arcsec\ radius circle is shown around the bulge of the galaxy and similar grey circles are placed around the galaxy for comparison. The position of HLX-1 is indicated with its 0\farcs3 error at 95\%.}
\label{fig:ch_im}
\end{figure}

No significant point source is detected inside the galaxy contours except HLX-1. 
We looked for an excess of counts in an extended 5\arcsec\ radius region encircling the bulge of the galaxy.
In the 0.3--8 keV energy band, we expect $5\pm2$ background counts in such a region and we may have in addition $5\pm2$ counts from HLX-1 due to the scattering of its emission on the detector (Figure~\ref{fig:ch_im}, lower panel). 
Finally, we expect a maximum of $10\pm3$ counts, while 19 counts were detected. 
We repeated the same estimation for different positions on the detector and found no other excess of counts (Figure~\ref{fig:ch_im}, upper panel).
We found a similar excess in the 0.5-6 keV energy range where the signal-to-noise is the best for ACIS-S: 15 counts detected, background of $3\pm2$ counts and HLX-1 emission of $3\pm2$ counts. 

The median energy from those photons is $0.92\pm0.10$~keV suggesting a soft emission. 
The emission from the bulge of spiral galaxies have previously been modeled by a \textit{mekal} model with temperatures of 0.3 to 0.6 keV \citep[e.g.][]{Humphrey:2004p1260}.
Assuming a \textit{mekal} model (at the redshift of the galaxy, $z=0.0224$, with solar metallicity and fixed absorption to the Galactic value of $1.8\times10^{20}$~atom~cm$^{-2}$), we find a temperature k$T=0.30\pm0.15$ and a flux of $\sim5\times10^{-15}$~\ergcm\ in the 0.2-10~keV range.
At the distance of the galaxy, this converts to an unabsorbed X-ray luminosity of $\sim6\times10^{39}$~\ergsec.
This level of luminosity is consistent with the integrated X-ray luminosity of spiral galaxies in general \citep[$10^{38}$--$10^{42}$~\ergsec,][]{Fabbiano:1989p3888}. 
This emission might contaminate \xmm\ and \swift\ HLX-1 data at low luminosities given their lower angular resolution.

\section{HLX-1 at low luminosities}
\label{low}

\subsection{\xmm\ spectrum XMM3}
\label{sec:xmmsp}

\begin{figure}
\centering
\includegraphics[width=\columnwidth]{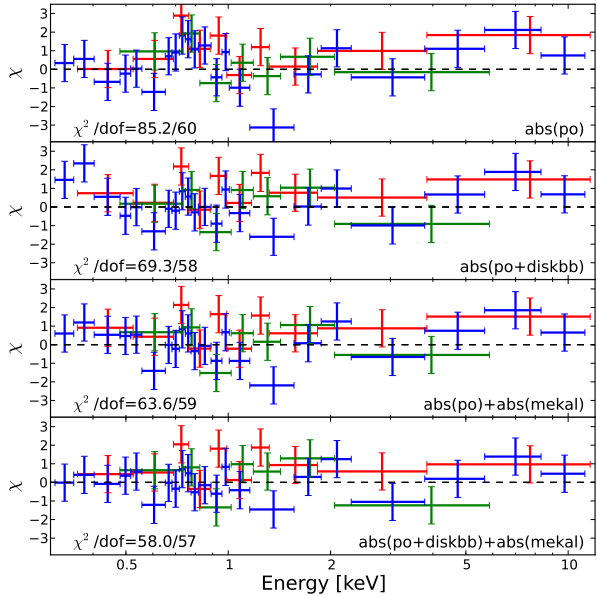}\\
\caption{XMM3 spectrum residuals for different models reported in Table~\ref{tbl:specfits}.}
\label{fig:xray_sp_xmm3_de}
\end{figure}

We fitted simultaneously the MOS1, MOS2 and pn spectra using Xspec and ignored channels with energy lower than 0.3 keV as well as channels tagged as bad. 
We used an absorbed power law and obtained a reduced $\chi^{2}$ of 1.42 with 60 degrees of freedom (Table~\ref{tbl:specfits}). 
The highest contributions to the $\chi^{2}$ are associated with a possible soft excess around 0.7~keV and a hard excess at high energies (Figure~\ref{fig:xray_sp_xmm3_de}, first panel).
We thus tested two possible additive components to the model to account for those features: (1) a \textit{diskbb} component to test for the presence of a thermal disk, and (2) a \textit{mekal} component which would correspond to a possible contamination from the galaxy bulge to the spectrum (see Section~\ref{sec:contamination}).

When adding a \textit{diskbb} component to the single absorbed power law model, the reduced $\chi^{2}$ is lowered to 1.20. However, a marked soft excess appears in the residuals and the hard excess is still present (Figure~\ref{fig:xray_sp_xmm3_de}, second panel).
The best fit is found for an $N_{\rm H}\sim6.5\pm1.5\times10^{21}$~atom~cm$^{-2}$, an order of magnitude higher than the XMM2 value and a lower temperature (k$T_{\rm in}\sim0.08$~keV) than for the \chandra\ data in the high state. The bolometric unabsorbed luminosity of the disk component is thus higher than in the high luminosity state (Section~\ref{high}).

The addition of an \textit{abs(mekal)} component --- at a fixed redshift of $z=0.0224$, with solar abundances and Galactic absorption as it could not be constrained --- to the single absorbed power law improved the reduced $\chi^{2}$ to an acceptable value of 1.08 and resulted in a lower photon index for the power law ($2.0\pm0.3$). 
The absorption of HLX-1 is not well constrained, so we fixed it to the XMM2 value which is the most precise estimate we obtained (see Table~\ref{tbl:specfits}). 
{We note that if the absorption is thawed, it tends to zero while the power law index further decreases.}
The residuals are shown in Figure~\ref{fig:xray_sp_xmm3_de} (third panel) and the folded spectrum is shown in Figure~\ref{fig:xray_sp_xmm}.
We found a temperature of $kT=0.43\pm0.09$~keV consistent with the expected emission from the galaxy (Section \ref{sec:contamination}). Moreover, the flux found for this absorbed \textit{mekal} component is $3.3\pm1.3\times10^{-15}$~\ergcm\ which is comparable to the estimate of the possible galaxy contribution found in the \chandra\ data (Section \ref{sec:contamination}).
We are thus confident that some emission from the galaxy contaminates the XMM3 spectrum of HLX-1, at a level of $\sim$17\% in flux at most.

Finally, we tested the presence of a \textit{diskbb} component in this last model. 
The resulting reduced $\chi^{2}$ is 1.02, the power law is harder ($1.6\pm0.4$) and the residuals seem featureless for this fit (Figure~\ref{fig:xray_sp_xmm3_de}, fourth panel). 
{We tested the significance of the \textit{diskbb} using the Bayesian posterior predictive probability values method. In the same way as in Section \ref{xmm12}, we generated 3000 simulated set of spectra based on the absorbed power law plus \textit{mekal} model and the XMM3 response files. We then computed the distribution of the $\chi^{2}$ improvement of the fit when adding a \textit{diskbb} component. The $\chi^{2}$ improvement of 5.6 for XMM3 translates to a significance of 93\% for the \textit{diskbb} component.}
We note that a \textit{diskbb} component is not needed to adequately describe the spectrum, but this possibility is intriguing and cannot be ruled out.



\subsection{Astrometry consistency}

\begin{figure}
\centering
\includegraphics[width=\columnwidth]{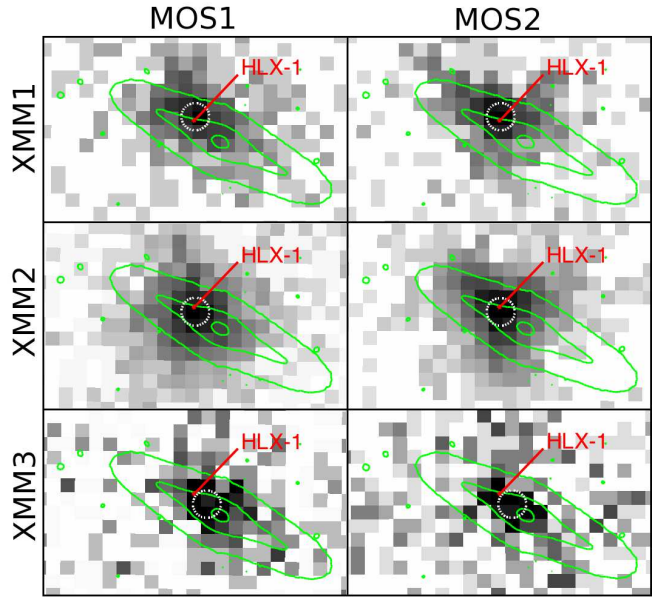}\\
\caption{\xmm\ MOS images of HLX-1 in the energy band 0.5--6 keV. The pixel size is 3\farcs2. The H-band galaxy contours are shown as in Figure~\ref{fig:ch_im}. The \xmm\ position error ($<$$3\farcs2$ at 95\%) is represented by dashed circles and the \chandra\ error circle ($0\farcs3$ at 95\%) is indicated by a red arrow.}
\label{fig:xmm_im}
\end{figure}

We aligned all images on the \chandra\ image, taking as a reference 3 point-like X-ray sources surrounding HLX-1. This allowed us to reduce the positional error by replacing the \xmm\ Absolute Measurement Accuracy\footnote{see \S4.7 in the \xmm\ Users Handbook, \url{http://xmm.esac.esa.int/external/xmm\_user\_support/documentation/ uhb/index.html}} of 4\arcsec\ by a residual error of less than 1\arcsec\ after alignment. We still have to take into account the Relative Pointing Error of 2\arcsec\ and the error on the centroid of the source, which yielded an uncertainty lower than the size of the pixels (3\farcs2) for all the images presented in Figure~\ref{fig:xmm_im}. We also note that the relative astrometry within all EPIC cameras is accurate to better than 1\farcs5. All errors are 95\% confidence.

We can expect that ESO 243--49 contributes to the X-ray emission of HLX-1 in XMM3, as determined using the higher angular resolution \chandra\ data (Section~\ref{sec:contamination}), since the bulge diffuse emission should be constant.
The source is slightly offset from the \chandra\ position which falls just outside the 95\% error circle. It is shifted towards the center of the galaxy ESO 243--49, indeed suggesting a contamination from the bulge of the galaxy.
However, the source detection task \textit{emldetect} did not split the source into two sources, indicating that the statistic is not sufficient to claim the presence of a contaminating source in the imaging data such as the galaxy bulge.
The PSF full width half maximum (FWHM) is 5\arcsec\ and 6\arcsec\ for MOS and pn, respectively, and the half energy width (HEW) is 14\arcsec\ and 15\arcsec. The distance between HLX-1 and the galaxy center is about 8\arcsec\ so the images cannot be used as evidence for contamination. 


\section{Hardness-intensity and hardness-rms diagrams}
\label{diagrams}

\subsection{\swift\ XRT spectra}

\begin{deluxetable*}{cccccccccc}
\tablecaption{Fit results for \swift\ XRT spectra \label{tbl:swift_sp}}
\tablewidth{0pt} 
\tablehead{\colhead{Name} & \colhead{Start} & \colhead{End} & \colhead{Hardness} & \colhead{Count rate} & \colhead{Unabs. $L_{\mathrm{X}}$} & \colhead{$\chi^{2}$/dof} & \colhead{C1} & \colhead{C2} & \colhead{$\chi^{2}$/dof} \\
 (1) & (2) & (3) & (4) &  (5) & (6) & (7) & (8) & (9) & (10)
}
\startdata
S1 & 54763 & 54784 & $0.34\pm0.04$ & $1.20\pm0.06$ & $6.4\pm0.6$ & 21.2/17 & $0.9\pm0.1$ & $1.3\pm0.4$ & 19.4/16 \\
S2 & 55048 & 55050 & $2.4\pm1.2$ & $0.15\pm0.09$ & $0.5\pm0.3$ & 30/35$^{*}$ & $<$$0.09$ & $0.4\pm0.3$ & 29.5/35$^{*}$ \\
S3 & 55059 & 55063 & $0.34\pm0.03$ & $3.30\pm0.13$ & $11\pm1$ & $14.4/21$ & \nodata & \nodata & \nodata \\
S4 & 55106 & 55164 & $0.27 \pm0.03$ & $1.86\pm0.09$ & $9.0\pm0.6$ & 26.6/19 & $1.4\pm0.2$ & $1.6\pm0.5$ & 25.5/18 \\
S5 & 55170 & 55225 & $0.34\pm0.05$ & $1.12\pm0.06$ & $5.8\pm0.6$ & 15.8/12 & $0.7\pm0.1$ & $1.6\pm0.5$ & 10.0/11 \\
S6 & 55237 & 55421 & $1.7\pm1.0$ & 0.$12\pm0.03$ & $0.2\pm0.1$ & 70/63$^{*}$ & 0 & $0.3^{+0.1}_{-0.2}$ & 65.4/63$^{*}$ \\
S7 & 55437 & 55453 & $0.21\pm0.02$ & $2.80\pm0.10$ & $11.8\pm0.3$ & 20.8/24 & \nodata & \nodata & \nodata \\
S8 & 55462 & 55490 & $0.21\pm0.03$ & $2.00\pm0.10$ & $9.6\pm0.6$ & 8.6/8 & \nodata & \nodata & \nodata \\
S9 & 55498 & 55543 & $0.26\pm0.04$ & $1.52\pm0.09$ & $8.3\pm0.6$ & 11.2/11 & $1.3\pm0.2$ & $1.5\pm0.6$ & 11.0/10 \\
S10 & 55548 & 55562 & $0.32\pm0.04$ & $0.8\pm0.3$ & $5.9\pm0.6$ & 12.6/14 & $0.84\pm0.14$ & $1.2\pm0.4$ & 10.8/13
\enddata 
\tablecomments{Columns: (1,2,3) Name, start and end day (MJD) of observation as reported in Figure~\ref{fig:swift_lc}; (4) Hardness is given as the ratio of fluxes $F[1-10~\mathrm{keV}]/F[0.3-1~\mathrm{keV}]]$; (5) Count rate ($10^{-2}$ count s$^{-1}$) in the band 0.3--10~keV; (6) Unabsorbed luminosity in $10^{41}$~\ergsec\ after fitting the XMM2 model multiplied by a constant; (7) $\chi^{2}$ and degrees of freedom (dof) for the fit; (8,9,10) Result of the fit with the two XMM2 components multiplied by factors C1 and C2, and goodness of the fit.\newline $^{*}$ Cash statistic was used and the number of PHA bins is indicated instead of dof.}
\end{deluxetable*}

\begin{figure}
\centering
\includegraphics[width=\columnwidth]{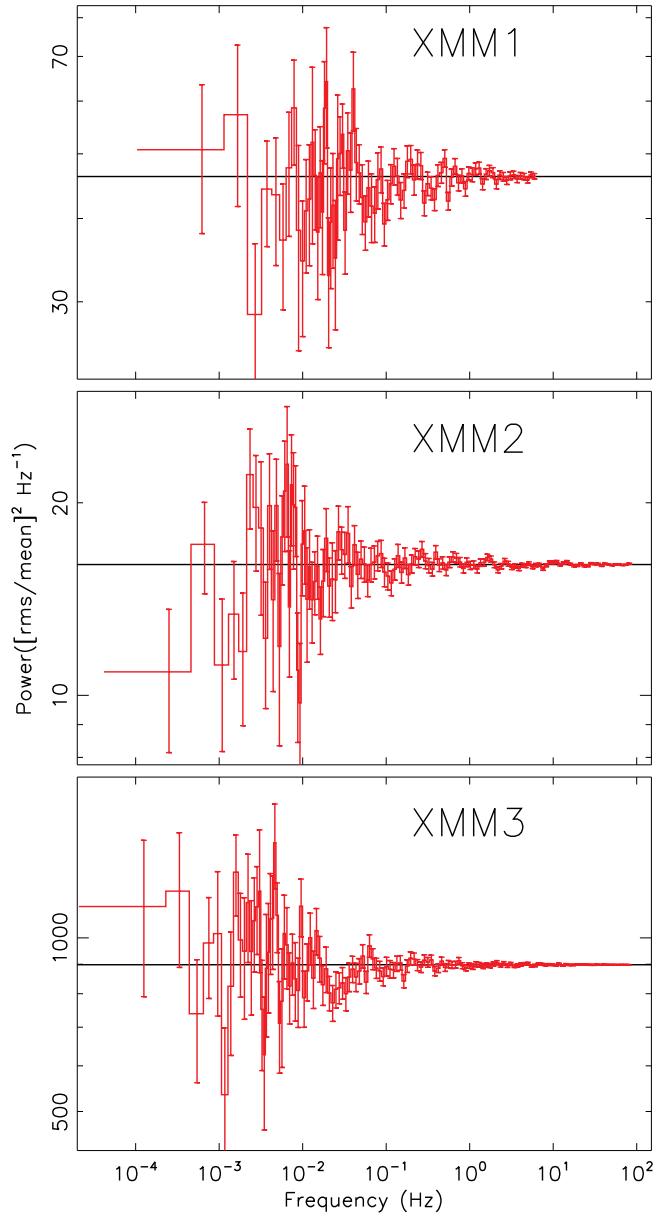}
\caption{The PDS of \xmm\ observations of HLX-1 using the pn camera. The black solid constant lines are the average PDS above 1 Hz, representing the Poisson level. \label{fig:lcfft}}
\end{figure}

We completed the study of the \chandra\ and \xmm\ spectra with the spectral fitting of grouped spectra from \swift\ XRT (Figure~\ref{fig:swift_lc}). We fitted each \swift\ XRT spectrum using Xspec. 
For all spectra, given the reduced statistical quality, we performed a basic fitting of the data using the XMM2 model (i.e. \textit{abs(diskbb+pow)}, see Table~\ref{tbl:specfits}) multiplied by a constant factor in order to estimate the unabsorbed luminosity.
{For each spectrum we estimated the hardness as the ratio of the count rates in the 1--10 and 0.3--1 keV bands. We then used WebPIMMS\footnote{http://heasarc.nasa.gov/Tools/w3pimms.html} to convert those ratios into flux ratios: in the low and high states (power law with photon index $\sim2.1$ and $\sim4.0$, respectively) we found a conversion factor of 1.5 and 0.9, respectively.}
We also used the following model: \textit{abs}(C1$\times$\textit{diskbb} + C2$\times$\textit{pow}) where C1 and C2 are two constants. The spectral parameters of the \textit{diskbb} and power law models were fixed at the values found in XMM2. The goal is to investigate the evolution in the contribution between the soft and hard components. 
All results are reported in Table~\ref{tbl:swift_sp}.

For the S3 and S7 spectra (high luminosity states, see Figure~\ref{fig:swift_lc}), we used a simple absorbed \textit{diskbb} model and reported the parameters of the fits in Table~\ref{tbl:specfits}.
There is no evidence in these spectra for the need of a hard energy component.
We note that the temperature for S7 is consistent with that derived from the  \chandra\ observation which was performed at a similar time. The normalization is lower for the \chandra\ spectrum, 
but the fluxes for both spectra remain consistent within the errors.

The basic spectral fitting suggests that the source is in the same state in S1, S4 and
S8. We merged the data from S1, S4 and S8 that approximately overlap in luminosity
assuming a $\sim$375 days periodicity of the source. In order to get a good fit, it is
necessary to add a hard component (power law) to the model. 
Results are reported in Table~\ref{tbl:specfits}.
The resulting parameters are consistent with those found for XMM2 at a similar level of luminosity.

Finally, we combined S2 and S6 to increase the signal-to-noise of the \swift\ XRT data in the low luminosity state. 
We obtained a good fit with an absorbed power law ($\Gamma=2.1\pm0.4$, Table~\ref{tbl:specfits}). We note that the flux is higher than for the XMM3 data. However, as the S2+S6 spectrum is from data obtained at different epochs and integrated over a long time, the comparison with XMM3 is not straightforward and differences in the two spectra are expected. The S2+S6 spectrum may also be affected by contamination from the galaxy bulge (Section~\ref{sec:contamination}).


\subsection{Timing analysis}
\label{time}


We calculated the power density spectra (PDS) for the light curves of HLX-1 from the three \xmm\ observations.
The light curves were split into four segments of equal lengths, and the PDS were calculated for each segment. For each light curve, all four PDS were merged and averaged by binning in frequency using a logarithmic factor 1.1, under the condition that each bin contains at least 20 individual PDS measurements. The errors were calculated from the sample standard deviation of PDS measurements in each bin.

The PDS of HLX-1 are shown in Figure~\ref{fig:lcfft}. The black solid lines denote the average PDS above 1~Hz. Those values deviate from the expected Poisson level by less than 0.4\%. We see that all PDS are flat, showing no significant intrinsic source variability, and no QPO. The fit results using a constant $C_{\rm P}$ are given in Table~\ref{tbl:timingfits}. To obtain a constraint on the variability, we followed the procedure adopted by \citet{Goad:2006p1221} and \citet{Heil:2009p1986}. In this procedure, the PDS are fitted with a broken power-law (BPL) model or a Lorentzian model plus a constant. 

For the BPL, the indices below and above the break frequency ($f_{\rm b}$) were assumed to be $-1$ and $-2$, respectively. The value $C_{1/f}$ corresponds to the power times the frequency below $f_{\rm b}$ from this model. The upper limits of $C_{1/f}$ for two values of $f_{\rm b}$ ($10^{-3}$ and 1~Hz) are shown in Table~\ref{tbl:timingfits}. We found that these limits are relatively independent of the assumed $f_{\rm b}$ value, at least within the $10^{-3}$--1~Hz range. Similarly, it is found to be fairly constant across AGN and GBHBs in the range 0.005--0.03 \citep{Papadakis:2004p1334}.
The upper limits of $C_{1/f}$ for XMM1 are consistent with the values generally observed for AGN/GBHBs in the thermal state. For XMM3, they are larger and thus not constraining. In XMM2, the upper limits of $C_{1/f}$ are low compared with typical values for AGN and GBHBs, in the range 0.005--0.03.
 
For the Lorentzian model, we assumed the quality factor to be 2
as in \citet{Goad:2006p1221} and \citet{Heil:2009p1986}. The upper limits of the integrated power of the Lorentzian $R^{2}$ are given in Table~\ref{tbl:timingfits} for two Lorentzian centroid frequencies: $10^{-3}$ and 1~Hz. They roughly increase with the centroid frequency assumed. 
Compared with the typical values of 0.01 for $R^2$ seen in the GBHB hard state \citep{vanderKlis:2006p1362}, the upper limits we obtained for HLX-1 are high for all three observations. 
Due to poor statistics, we thus cannot place any firm constraints on the spectral state using the timing analysis.

\begin{deluxetable*}{cccccccc}
\tablecaption{Results of the PDS Fitting \label{tbl:timingfits}}
\tablewidth{0pt}
\tablehead{\colhead{Obs} & \colhead{$C_{\rm P}$} & \colhead{$\chi^2_\nu$ ($\nu$)} & \multicolumn{2}{c}{$C_{1/f}$ ($10^{-4}$, BPL)} & \multicolumn{2}{c}{$R^{2}$ ($10^{-2}$, BLN)} & \colhead{${\rm rms} (\%)$} \\
 & & & $10^{-3}$ Hz & 1 Hz & $10^{-3}$ Hz & 1 Hz \\
 (1) & (2) & (3) & (4) &  (5) & (6) & (7) & (8)
}
\startdata
XMM1 &   46.11$\pm$0.22 & 1.13(83) & $<$98.8 & $<$58.3&  $<$4.2 & $<$326.3 & $<$$32.3$  \\
XMM2 &   16.02$\pm$0.01 & 1.28(120) & $<$2.7 & $<$2.5 & $<$0.49 & $<$58.8 &$8.4^{+9.1}_{-8.4}$\\
XMM3 & 898.11$\pm$0.55 & 0.79(127)& $<$646.8 & $<$579.9 & $<$79.2 & $<$1293.7 & $<$96.7
\enddata 
\tablecomments{Columns: (1) Observations; (2) Best-fitting constant Poisson level in [rms/mean]$^{2}$~Hz$^{-1}$; (3) Reduced $\chi^2$ and degrees of freedom of the fits with a constant Poisson level; (4) Best-fitting $C_{1/f}$ assuming a break frequency of $10^{-3}$ Hz; (5) Best-fitting $C_{1/f}$ assuming a break frequency of 1 Hz; (6) Best-fitting $R^2$ assuming a Lorentzian centroid frequency of $10^{-3}$ Hz; (7) Best-fitting $R^2$ assuming a Lorentzian centroid frequency of 1 Hz; (8) Fractional rms variability within 0.0001--0.1 Hz, assuming the Poisson level to be the average PDS above 1 Hz. All limits are at a 90\% confidence level, except for the fractional rms (column 8), which is 1$\sigma$.}
\end{deluxetable*}

\subsection{Combined diagrams}



\begin{figure}
\centering
\includegraphics[width=\columnwidth]{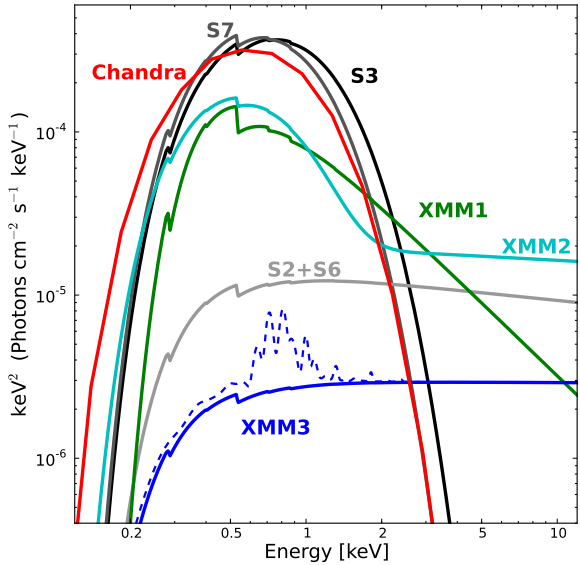}\\
\caption{Models of X-ray spectra of HLX-1 in different states as obtained from the best fit to the data (see Table~\ref{tbl:specfits}). For XMM3, the \textit{abs(pow)+abs(mekal)} model is shown as a dashed line, and the model without the \textit{mekal} component as a thick line. For Chandra, the effect of pile-up has been corrected.}
\label{fig:xray_spmo}
\end{figure}

\begin{figure}
\centering
\includegraphics[width=\columnwidth]{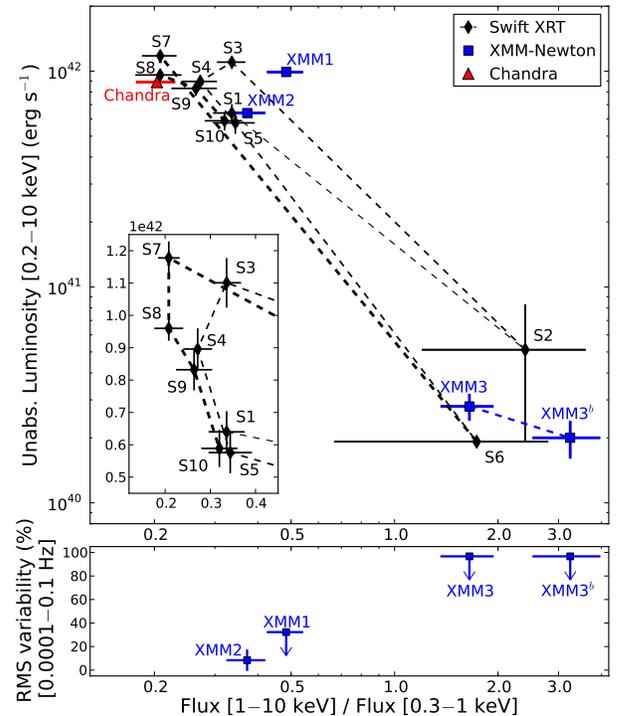}\\
\caption{Hardness-intensity (top) and hardness-rms diagrams (bottom) for HLX-1. This plot includes data from \swift, \xmm\ and \chandra. The hardness ratios and luminosities are model dependent, and were obtained from the best fit of the spectra or using a conversion factor for the \swift\ XRT. XMM3$^{b}$ corresponds to the model with the \textit{mekal} component removed (see Section~\ref{low}). We show part of the HID with linear axes as an inset to better follow the evolution of the source in the high state.}
\label{fig:xray_hid_hrd}
\end{figure}

We show the best fit models for all the fitted spectra in Figure~\ref{fig:xray_spmo}, which
illustrates the evolution of the soft disk emission and the hard tail during spectral state transitions.
For each \swift, \xmm\ and \chandra\ spectra, we estimated the unabsorbed luminosity in the range 0.2--10 keV for the best fit, and calculated the flux ratio between the bands 0.3--1 and 1--10 keV. Those energy ranges were also used by \citet{Godet:2009p1106}, so the HID shown in Figure~\ref{fig:xray_hid_hrd} is an updated version of their Figure~2.

We plotted in this diagram values coming from different instruments with independent calibrations. There could be systematic errors in the different calibrations, leading to a discrepancy of up to 20\% in the normalization of the flux as it could be observed during simultaneous observations \citep[e.g.][]{Tsujimoto:2011p1781}.
We indeed observed a difference between the \swift\ XRT S7 and \chandra\ ACIS-S spectra of HLX-1 (Figure~\ref{fig:xray_spmo}) taken over a similar period. However, for these two spectra, hardness ratio and luminosity estimates are consistent within this range of errors. 
Consistent trends are seen in both the \swift\ and \xmm\ data, we are thus confident that the spectral variations detected in this work are real and significant.

We also computed variability rms values and limits in the 0.0001--0.1~Hz range. They are reported in  Table~\ref{tbl:timingfits} and plotted on the HRD in Figure~\ref{fig:xray_hid_hrd}.

\section{Discussion}
\label{discuss}

\subsection{Spectral states and transitions}

The high luminosity state of HLX-1 is associated with a soft emission with a dominant thermal disk emission, while the low luminosity state is found to be significantly harder and power-law-like. 
This is reminiscent of the behavior of GBHBs \citep[e.g.][]{Remillard:2006p561}, and the HID and HRD shown in Figure~\ref{fig:xray_hid_hrd} are consistent with similar diagrams for GBHBs \citep[e.g.][]{Belloni:2010p1608}.
HLX-1 therefore clearly showed transitions from one state to another, confirming the first evidence found by \citet{Godet:2009p1106}.
However, fewer luminosity states have been observed and therefore the diagrams are not equally sampled than for GBHBs.

The high luminosity state of HLX-1 is well described by an optically thick disk model with temperatures of 0.18 to 0.26 keV (Table~\ref{tbl:specfits}). Moreover, the low variability level in the XMM2 data ($\sim$10\%) can reasonably be ascribed to the dominance of the thermal disk component. All this is consistent with HLX-1 being in a thermal state.
The temperature of the thermal component is 4 to 5 times lower than for typical GBHBs with stellar-mass black holes \citep[$\sim1$~keV, e.g.][]{Remillard:2006p561}. 
If this soft component we measured is due to emission from the inner region of an accretion disk, and the disk extends close to the ISCO, then the temperature is related to the mass of the black hole by the relation $T_{\rm in} \propto M^{-1/4}_{\rm BH}$ \citep{Makishima:2000p5954}. 
It is thus possible that the lower temperature is due to the presence of a black hole with a higher mass --- by a factor of $\sim200$ to 500 --- than a typical stellar-mass black hole ($\sim10$~\msun), leading to a possible mass of few $10^{3}$~\msun. 
This argument has already been used by \citet{Miller:2004p1567} for 6 bright ULXs ($>10^{40}$~\ergsec) to strengthen their classification as IMBH candidates even if the thermal component was not dominant. In our observations of HLX-1 in the thermal state, this component is clearly dominant.
A complementary study of HLX-1 in the thermal state by \citet{Davis:2011p4901} using the relativistic disk model BHSPEC provided robust lower and upper limits with $3000 < M < 3\times10^{5}$~\msun, firmly placing HLX-1 in the IMBH regime. Godet et al. (2011, in preparation) tested additional models to physically constrain the nature of HLX-1 and they also found mass estimates in the IMBH range.

\begin{figure}
\centering
\includegraphics[width=\columnwidth]{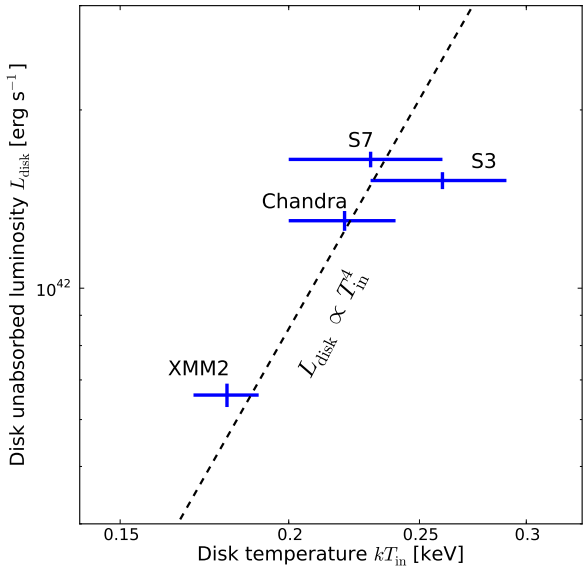}\\
\caption{Disk bolometric unabsorbed luminosity as a function of disk temperature $kT_{\mathrm{in}}$ (see values in Table~\ref{tbl:specfits}). The best fit with a fixed $L_{\mathrm{disk}} \propto T_{\mathrm{in}}^{4}$ is indicated with a dashed line. The XMM3 point should be taken with caution as the disk component cannot be confirmed nor ruled out in the X-ray spectrum.}
\label{fig:LvsT}
\end{figure}

We report in Figure~\ref{fig:LvsT} the temperature and bolometric luminosity of the disk in the thermal state from the S3, S7, \chandra, XMM2 observations (see also Godet et al. 2011, in preparation). 
All those measurements in the thermal state are consistent with the correlation between luminosity and temperature expected from a geometrically thin, optically thick accretion disk \citep[$L \propto T^{4}$,][]{Shakura:1973p1981}. 
This supports the idea of HLX-1 being in a generic thermal state as this relation is also observed for GBHBs in this state \citep[e.g.][]{Kubota:2004p5432,Remillard:2006p561}. It is however different to the ULX branch reported e.g. by \cite{Soria:2007p5488}, where an anti-correlation between disk luminosity and temperature is seen for two ULXs.

During the XMM3 observation at low luminosities, the spectrum of HLX-1 is mainly described by a power law model with a photon index that falls in the range 1.4 to 2.1 (Table~\ref{tbl:specfits}), indicating a hard state as observed for GBHBs (we presented preliminary results in \citealt{Farrell:2011p2140}).
We could not constrain the rms variability of the source during the XMM3 observation due to low statistics. Therefore, a high rms value of $\sim$30\%, as observed for GBHBs in the hard state \citep[e.g.][]{Belloni:2010p1608}, cannot be ruled out and would be consistent with our data.
Assuming that this hard state is similar to the one observed for GBHBs, one would expect enhanced radio emission \citep{Fender:2004p3226}. Such a detection would further confirm this state as the canonical hard state of GBHBs and would allow a mass estimate using the fundamental plane \citep{Merloni:2003p2145}.

{In the XMM3 spectrum, dominated by a power law component, we found marginal evidence for a soft disk component with temperature $kT_{\rm in}=0.07\pm0.04$~keV, with a significance of 93\%. Such a disk component would be consistent with the $L \propto T^{4}$ relation reported in Figure~\ref{fig:LvsT}, as the expected temperature for a disk unabsorbed luminosity of $4\times10^{40}$~\ergsec\ is 0.09~keV. A disk could thus be either extending down to the ISCO, or maybe truncated as the best fit temperature value is lower than the temperature expected from the $L \propto T^{4}$ relation. This would again suggest a resemblance with GBHBs.}

During the XMM1 observation of HLX-1, the spectrum is found to be well fitted with a steep power law of photon index $\sim$3.5. Such a spectral state is also observed for some GBHBs \citep[steep power law state, e.g.][]{Remillard:2006p561}, supporting the similarities between HLX-1 and GBHBs.



The HLX-1 long term variability seen in Figure~\ref{fig:swift_lc} presents similarities with some lightcurves of GBHB \citep[e.g.][]{Gierlinski:2006p1977,Done:2007p1217} and more generally of soft X-ray transients when they show outbursts \citep{King:1998p1901}. We observed a fast rise ($\sim$10 days or less), followed by an exponential decay (100 to 200 days) of the X-ray flux from HLX-1 (FRED-like outburst). 
This could be explained by theoretical models of the hydrogen ionization instability controlling the fast rise \citep{King:1998p1901}. 
The cause of this variability has been investigated in more details by \citet{Lasota:2011p4895}.
They concluded that HLX-1 is unlikely to be explained by a model in which outbursts are triggered by thermal-viscous instabilities in an accretion disc \citep{Lasota:2001p5099}, and they argue that a more likely explanation is a modulated mass-transfer due to tidal stripping of a star in an eccentric orbit around the IMBH. This would still explain a change in the accretion rate, leading to the state transitions we observe, and might also explain the rarity of a source like HLX-1, which shows unique properties for a ULX.

\subsection{Comparison with other ULXs}


It was suggested that most ULXs (few $10^{39}$ to $10^{41}$~\ergsec, with power law spectra) are stellar-mass systems accreting at Eddington ratios of the order of 10--30, with mild beaming factors $b\gtrsim0.1$ \citep{King:2009p2048}. 
If a supercritical accretion disk --- as proposed in the Galactic system SS433 for example --- is seen face-on, the expected luminosity could indeed reach $10^{41}$~\ergsec\ \citep{Fabrika:2007p4904}.
In the case of HLX-1 in the highest state, the Eddington ratio would reach 170 and the beaming factor $b\sim2.5\times10^{-3}$ \citep{King:2011p4959}.
\cite{Freeland:2006p4898} calculated the broad-band radio--X-ray spectra predicted by micro-blazar and micro-quasar models for ULXs. They argued that a disk and a jet could be present in the system in a high/hard state close to the Eddington luminosity with a high accretion rate and gravitational energy released all the way down to the ISCO, leading to a hard spectrum ($\Gamma$=1.4--2.1) at high observed luminosities ($>4\times10^{39}$~\ergsec). 
They also proposed the existence of milli-blazars (IMBH and beamed emission) with luminosities $>10^{41}$~\ergsec. The spectrum of HLX-1 in the low luminosity state might be explained by such a model. However, the transition to a thermal state at even higher luminosities is not consistent with this picture. The increase in luminosity and softening of the source (decrease of the peak energy as can be seen in Figure~\ref{fig:xray_spmo}) are not consistent with relativistic beaming variations which would have an opposite effect \citep[e.g.][]{King:2009p2048}. All this indicates that the spectral variability we observe for HLX-1 is not due to beaming variations.





\citet{Gladstone:2009p2047} showed that simple spectral models commonly used for the analysis and interpretation of some ULXs (power law continuum and multicolor disk black body models) are inadequate  for a small sample (12 ULXs) of nearby, low luminosity ULXs that have high quality spectra. 
Two near ubiquitous features are found in the spectra: a soft excess and a rollover in the spectrum at energies above 3~keV. They suggested the existence of a new ultra-luminous state with super-Eddington accretion flows. This would favor the presence of stellar-mass black holes rather than IMBHs for some ULXs.
In the spectra of HLX-1, which reaches luminosities two orders of magnitude higher than the average ULX, we found no evidence of such a break above 3~keV. However, the number of counts, even in XMM2, is probably insufficient to detect this feature (\citealt{Gladstone:2009p2047} claim the need of 10\,000 counts).
Godet et al. (2011, in preparation) discuss in more details the comparison with ULXs claimed to be in a ultra-luminous state, but the variety of X-ray spectra we observe for HLX-1 in this work, all in the ultra-luminous range, is clearly unique for a ULX. In particular, the thermal state observed at high luminosities (therefore probably at the highest accretion rate) for HLX-1 is clearly distinct to the proposed ultra-luminous state spectrum for some ULXs.


Finally, the X-ray observations reported in this work indicate that HLX-1 more resembles GBHBs scaled to higher luminosities (three orders of magnitude) than any other category of ULXs.
In this picture, the bolometric luminosity (hence the X-ray luminosity) of the source in the thermal state should be lower than the Eddington luminosity for the system, giving a limit on the mass of the black hole of $>$9\,000~\msun.
The unabsorbed luminosity in the hard state is 2.2\% of the highest luminosity recorded in this work for which we assume a sub-Eddington luminosity (see Table~\ref{tbl:specfits}). This level of luminosity is consistent with the fraction of Eddington luminosity observed for GBHBs in the hard state \citep[$<$2\%, e.g.][]{Maccarone:2003p5296,Gierlinski:2006p1977}, further supporting the analogy of HLX-1 with GBHBs.
It is generally observed for GBHBs that in the hard state and the steep power law state, the X-ray luminosity is 1\% and 100\% of the Eddington luminosity \citep{Remillard:2006p561}, respectively, leading to complementary mass estimates of 16\,000 and 20\,000~\msun.

\section{Conclusion}

We observed the bright ULX ESO 243--49 HLX-1 with different X-ray observatories at luminosities ranging from $2.1\times10^{40}$ to $1.25\times10^{42}$~\ergsec.
The spectral analysis revealed that HLX-1 showed three states: a thermal state at high luminosity, a hard state at low luminosities, and a steep power law state.
These states are typically observed for GBHBs, which have luminosities at least three orders of magnitude lower than HLX-1.
The clear spectral changes we observed are inconsistent with models of milli-blazars where the emission is strongly beamed toward us, and the spectra of HLX-1 does not resemble spectra of ULXs proposed to be in an ultra-luminous state with super-Eddington accretion.
Instead, the source is much better explained as being homologous to GBHBs, which have non-beamed and sub-Eddington emission.
The relatively low temperature of HLX-1 in the thermal state suggests the presence of an IMBH of few $10^{3}$~\msun. Finally, in this picture, the source luminosity should be lower than the Eddington luminosity, which converts to a lower limit on the mass of the black hole of $>$9000 $M_\odot$, placing HLX-1 in the IMBH range.


\acknowledgments

We are grateful for the comments of the referee which helped us improve and strengthen the paper.
We thank Harvey Tananbaum and the Chandra team for approving the Chandra DDT observations.
MS thanks Josh Grindlay and Jeff McClintock for helpful discussions.
MS acknowledges supports from NASA/Chandra grants GO0-11063X, DD0-11050X and NSF grant AST-0909073. 
SAF acknowledges funding from the UK Science and Technology Funding Council and the Australian Research Council. SAF is the recipient of an Australian Research Council Post Doctoral Fellowship, funded by grant DP110102889.
Based on observations from XMM-Newton, an ESA science mission with instruments and contributions directly funded by ESA Member States and NASA.
This research has made use of data obtained from the Chandra Data Archive and software provided by the Chandra X-ray Center (CXC) in the application packages CIAO, ChIPS, and Sherpa.
This work made use of data supplied by the UK Swift Science Data Centre at the University of Leicester.

{\it Facilities:} \facility{XMM}, \facility{CXO}, \facility{Swift (XRT)}

\bibliographystyle{apj} 
\bibliography{../../../ref.bib}

%
%
%
%
%
%

\end{document}